## Very thin (111) NiO epitaxial films grown on c-sapphire substrates by pulsed laser deposition technique

**Santosh Kumar Yadav and Subhabrata Dhar**[a)]

*Department of Physics, Indian Institute of Technology Bombay, Mumbai 400076, India*

[a]*Electronic mail: dhar@phy.iitb.ac.in*

(111)NiO epitaxial layers are grown on c-sapphire substrates by pulsed laser deposition (PLD) technique. Structural and morphological properties of the films are studied using in-plane as well as out-of-plane high resolution X-ray diffraction and atomic force microscopy techniques as functions of growth temperature, oxygen pressure and the pulses count of the laser. The study shows that continuous epitaxial films of thickness as low as 3 nm with high crystalline quality, smooth surface and interface morphology can be grown by this technique. The study also reveals the co-existence of 60°-rotated (111) triangular domains of NiO in the film. The study also evidences the presence of a very low density of 60-degree dislocations in these films. Density of screw and edge dislocations are also estimated to be quite low. It has been found that growth-temperature, oxygen partial pressure and the film thickness can influence differently the density of various dislocation types. These parameters are also found to affect significantly the strain developed in the films.

### I. INTRODUCTION

Nickel oxide (NiO) is a wide-bandgap semiconductor, whose bandgap is reported to range between 3.6 to 4 eV.[1,2] The material, which has a rocksalt lattice structure with excellent stability in air, is antiferromagnetic in nature with Néel temperature of 525K.[3,4] These properties have made NiO a potential candidate for various devices such as exchange biased systems[5], spin-valves[6,7], UV photo-detectors[8] and UV-light-emitting diodes (LEDs)[9]. NiO is one of the few oxide semiconductors, which can exhibit a stable p-type conductivity.[2,10,11] This has led to its application as hole transport layers in organic solar cells[12,13] and p-type transparent conducting oxide (TCO) films[14]. It has been found that stable p-type doping in many wide bandgap oxide semiconductors such as ZnO, $In_2O_3$ and $Ga_2O_3$, which have tremendous potential for UV optoelectronics, is a big challenge. During growth, these materials are often unintentionally doped n-type.[15–18] This makes it difficult to develop p-n junction based optoelectronic devices out of these oxide semiconductors even though they have superior optoelectronic properties. This gap can be bridged by growing p-n heterojunctions of these oxides with p-NiO.[19]

Polycrystalline NiO films grown on polymer or glass substrates are widely used for device purpose. Such layers often exhibit p-type conductivity, which is attributed to nickel vacancies and/or oxygen interstitials as these defects are expected to act as shallow acceptors in NiO.[20,21] However, it should be noted that polycrystallinity results in a large density of structural defects in the film, which adversely affects the transport and optical properties of the material. Improvement in the performance of the devices thus requires crystalline films. This demand, in recent years, has generated a lot of interests in growth of epitaxial NiO films, which is also motivated by the desire to understand the basic properties of the material. There are reports of epitaxial growth of NiO on various substrates like sapphire, cubic yttria-stabilized zirconia (c-YSZ) and MgO, using different growth techniques such as pulsed laser deposition (PLD)[14,22,23], magnetron sputtering[24] and mist chemical vapor deposition (CVD)[10], chemical solution[25] and molecular beam epitaxy (MBE)[26] techniques. Many of these efforts have led to the growth of films with good epitaxial quality and smooth morphology. However, a thorough study on the structural defects such as misfit induced dislocations as well as the influence of various growth parameters on the structural and electronic properties of the film are still lacking. These studies are highly important for the enhancement of performance of the devices based on this material.

Another important aspect is to explore the minimum thickness, down to which the layer remains to be smooth and continuous. Note that an epitaxial film of NiO with thickness of a few nanometres could be an interesting candidate for tunnelling barrier in spintronic devices.[27]

Here, we report the growth of [111] oriented NiO epitaxial layers on c-sapphire substrates using PLD technique. A systematic study of growth, structural and morphological properties of the films is carried out as functions of the growth temperature, oxygen partial pressure and the layer thickness. Our study explores how the strain in the epitaxial layer and the densities of various types of misfit dislocations vary with those parameters. The study further shows that continuous epitaxial films with high crystalline quality and thickness as low as 3 nm can be grown by this technique at optimized growth conditions.

## II. EXPERIMENTAL DETAILS

[111] oriented NiO epitaxial films were deposited on c-sapphire substrates. A KrF excimer laser with wavelength of 248 nm and pulse width of 25 ns was used to ablate the NiO pellet. Energy density of the laser pulse was kept at 1.2 $Jcm^{-2}$ at a frequency of 5 Hz. All substrates were cleaned by sonication in an ultrasonic bath with trichloroethylene, acetone and methanol for 3 minutes each. Substrates were then dipped in 1:20 solution of HF in $H_2O$ for one minute followed by rinsing in methanol and drying under nitrogen flow. Base pressure of the chamber was measured to be less than $1\times10^{-5}$ mbar. Several samples were grown under various oxygen partial pressures ranging from $1\times10^{-1}$ to $3\times10^{-4}$ mbar, at various growth temperatures ranging from 200 to 700°C, and for different number of laser pulses ranging from 500 to 15000. After switching off the laser, the sample was kept at the same temperature for 30 more minutes before it was cooled down. Partial pressure of oxygen inside the chamber was maintained throughout the growth as well as during cooling down stage. Samples were removed from the chamber at room temperature. Both in-plane and out-of-plane x-ray diffraction studies were carried out on these samples using a Rigaku SmartLab HRXRD system. Thickness of these layers is examined with cross-sectional scanning electron microscopy (SEM) and x-ray reflectivity (XRR) studies. XRR profiles are analyzed by “GlobalFit” software supplied by Rigaku.[28] Surface morphology of these films was investigated by atomic force microscopy (AFM) and SEM techniques. All samples investigated in this study are found to be highly resistive (~ GΩ).

## III. RESULTS AND DISCUSSION

### A. $O_2$ partial pressure

A set of samples are grown at different oxygen partial pressures ($P_{O_2}$) ranging from $3 \times 10^{-4}$ to $1 \times 10^{-1}$ mbar at a fixed growth temperature of 600°C and for 10000 laser pulses. This series is termed as $P_{O_2}$- series.

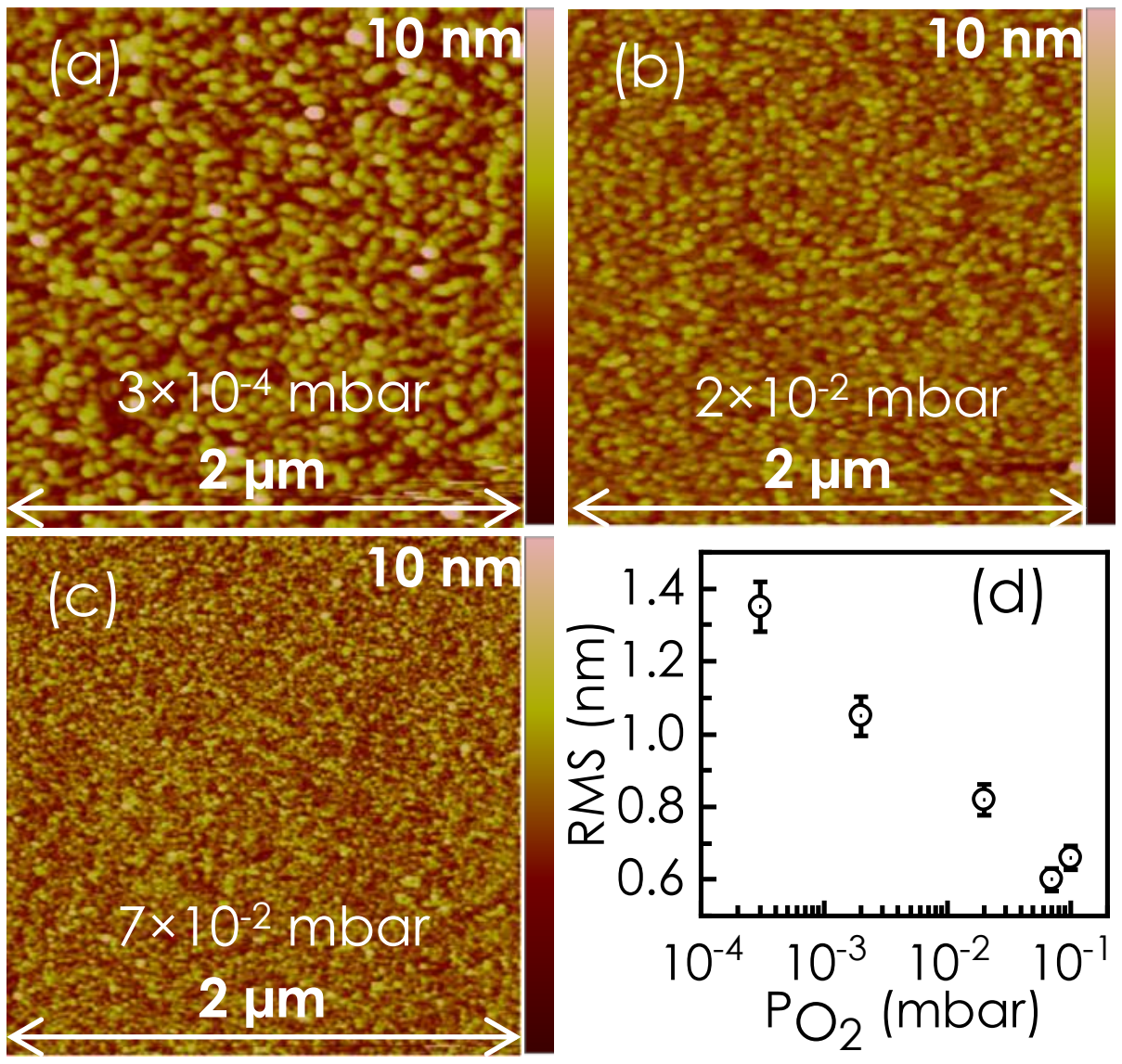

Fig. 1 (a-c) AFM top view images of samples grown at different $P_{O_2}$. (b) RMS roughness as a function of $P_{O_2}$.

Figure 1(a-c) shows the top-view AFM image of the sample grown at various oxygen partial pressures keeping the growth temperature constant at 600°C. All samples of this series are found to have continuous morphology with RMS-roughness ranging between 0.6 to 1.4 nm. The root mean square (RMS) surface roughness as obtained from the AFM scans is plotted as a function of oxygen pressure in Fig. 1(d). Evidently, the roughness decreases with the increase of $P_{O_2}$.

X-ray $\omega - 2\theta$ profiles for films grown with different oxygen partial pressures are shown in figure 2(a). As can be seen from the figure, (111) and its higher order reflections are the only visible features associated with NiO. This finding clearly shows [111] directional growth of these films. Interestingly, at low oxygen partial pressures, an additional peak appears at 44.54°. This can be identified as (111) reflection of face centered cubic (fcc) phase of Nickel crystals. Moreover, the absence of other reflections associated with fcc phase of Ni indicates oriented inclusion of Ni clusters in the NiO lattice when the layers are grown under low partial pressure of oxygen. A detail study on the inclusion of oriented Ni clusters in NiO epitaxial films will be the subject matter of a forthcoming publication. A weak feature that appears at 64.5° can be attributed to $(20\bar{2}5)$ reflection of sapphire substrate arising due to misoriented grains.

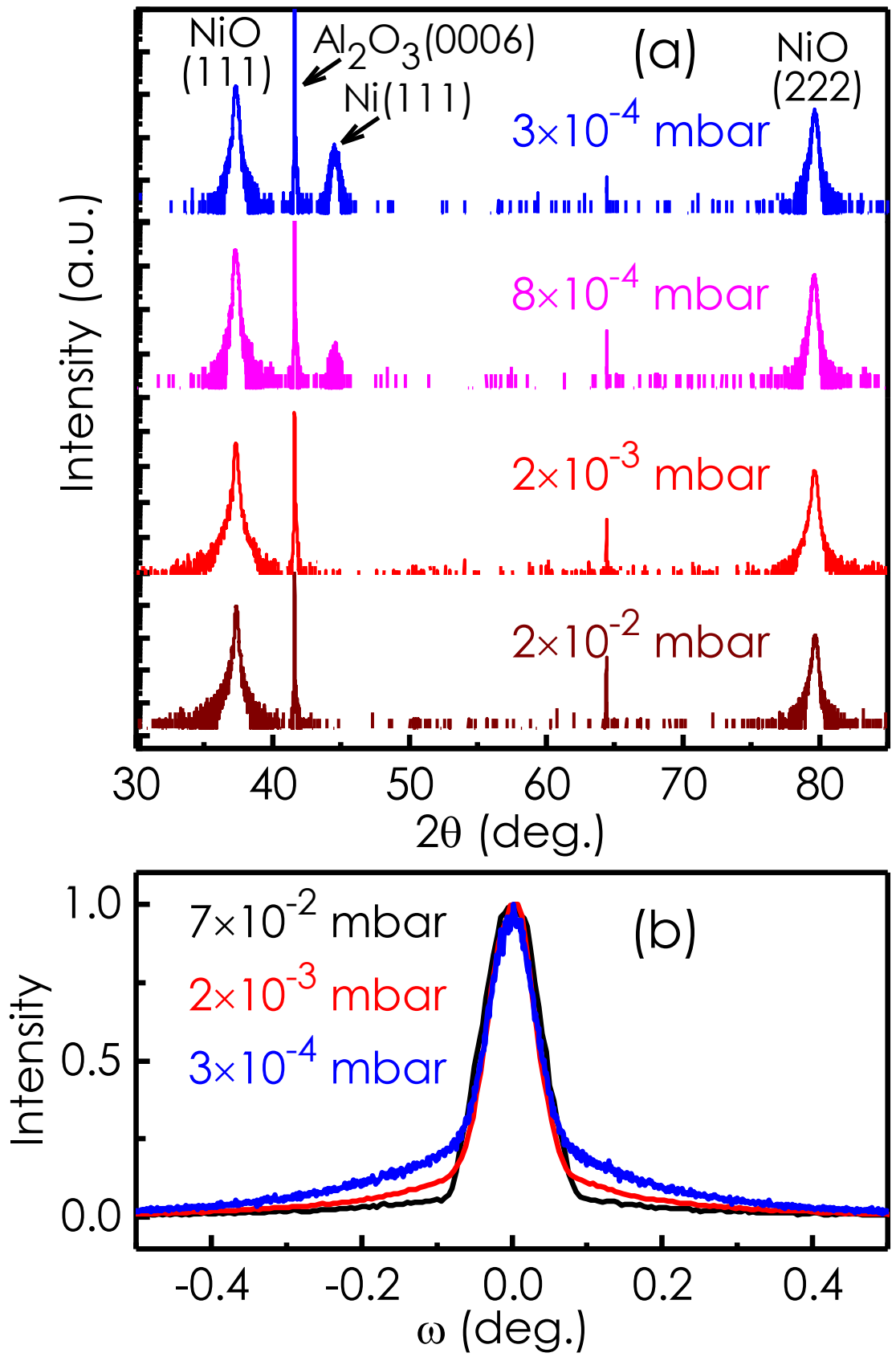

Fig. 2 (a) $\omega - 2\theta$ scans for samples grown at different oxygen partial pressures. (b) Rocking curves for (111)-NiO reflection.

Figure 2(b) compares the rocking curves ($\omega -$ scans) associated with (111)-NiO peak for samples grown with different oxygen partial pressures. Barring the tail parts, the features are quite identical for these samples with a full width at half maximum (FWHM) value of only ~0.08°. This suggests a low density of screw type dislocations in these layers. This width corresponds to screw dislocation density of $3.4 \times 10^8$ cm$^{-2}$.[29,30] However, the tail parts show different amounts of broadening for these samples. It is noticeable that the tail broadening for these samples increases with the decrease of the partial pressure of oxygen.

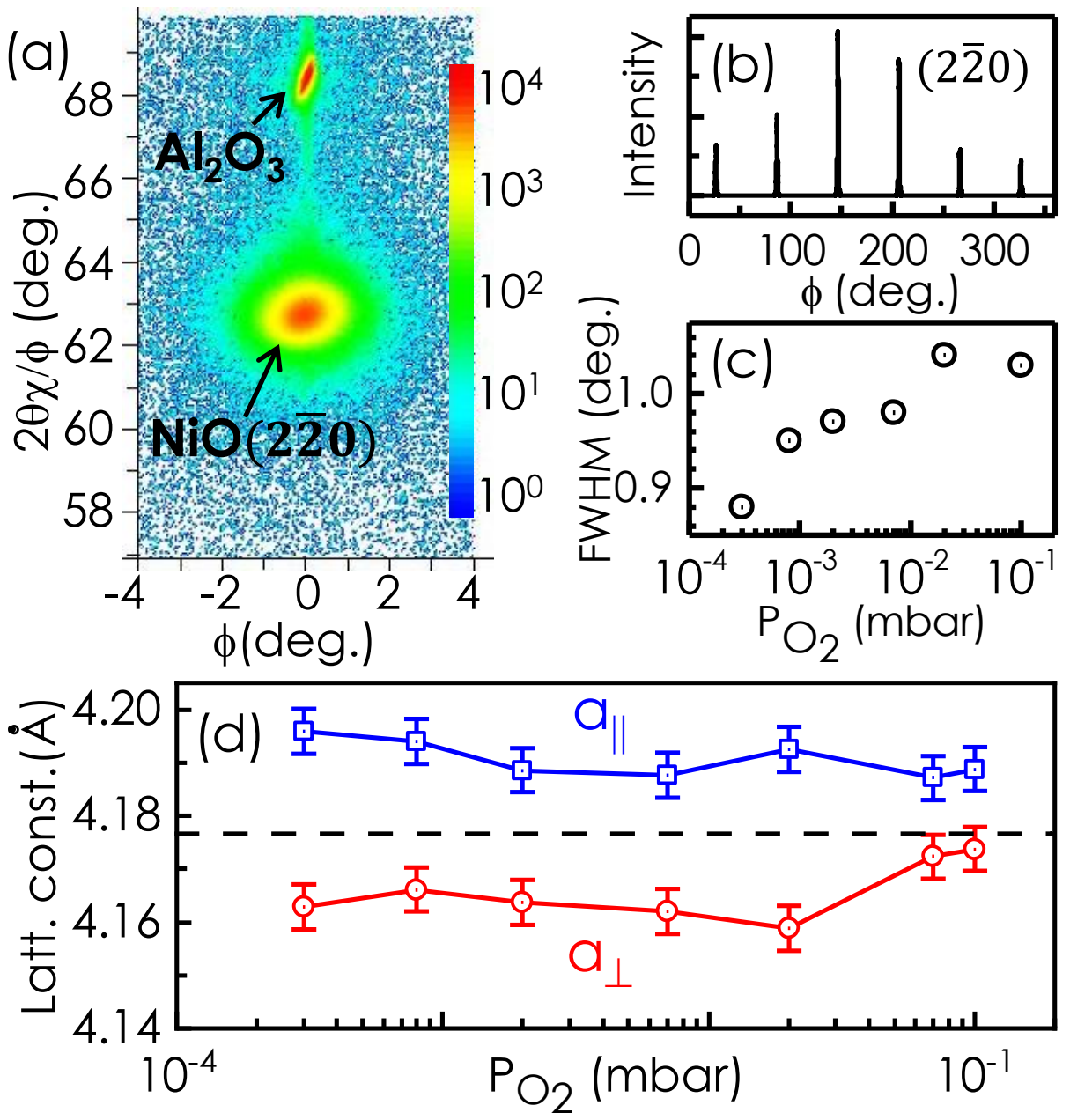

Fig. 3(a) In-plane reciprocal space map for the sample grown with $P_{O_2} = 2 \times 10^{-2}$ mbar. (b) Wide angle $\phi$-scan for (2$\bar{2}$0) reflection for the same sample. (c) FWHM of the in-plane $\phi$-scan peaks as a function of $P_{O_2}$. Error in FWHM estimation is ~ 0.1%. (d) Lattice constants $a_\parallel$ and $a_\perp$ as a function of $P_{O_2}$, dashed line represents the lattice constant of bulk NiO.

Figure 3(a) shows the in-plane reciprocal space map (RSM) for the sample grown with $2\times10^{-2}$ mbar oxygen pressure. NiO (2$\bar{2}$0) and sapphire (30$\bar{3}$0) features are clearly visible. Panel (b) shows the (2$\bar{2}$0) $\phi$-scan recorded for the same sample in a wide angular range. Six equidistant peaks are clearly visible. Similar $\phi$-scans are obtained for all samples investigated here. It should be noted that the cubic NiO lattice is expected to show three-fold symmetry about [111] direction. Appearance of six equally spaced peaks thus suggests the co-existence of 60°-rotated triangular domains.[24] It has been shown that the growth of such domains can indeed be possible on terraces, which are separated by 0.22 nm high atomic steps, on the c-surface of sapphire.[31] Observations of Fig. 2, 3(a) and 3(b) thus confirm the epitaxial nature of these layers. Figure 3(c) plots the FWHM of the (2$\bar{2}$0) $\phi$-scan as a function of the partial pressure of oxygen. Evidently, FWHM for these in-plane rocking curves increases with $P_{O_2}$, suggesting an increase in edge dislocation density with oxygen partial pressure. Note that the broadening of the in-plane rocking curves is an order of magnitude larger than that of its out-of-plane counterparts shown in Fig. 2(b). This suggests that the density of edge dislocations is much higher than that of screw dislocations in these layers. In fact, the lowest edge dislocation density is estimated to be $6.5 \times 10^{10}$ cm$^{-2}$,[29,30] which is more than two orders of magnitude larger than that of screw dislocations found in these samples.

Figure 3(d) plots the variation of the in-plane lattice constant $a_\parallel$ estimated from in-plane RSM images [as shown in Fig. 3(a)] and $a_\perp$estimated from $\omega - 2\theta$ scans [as shown in Fig. 2(a)] as a function of $P_{O_2}$. For all the samples in the series, $a_\parallel$ is found to be larger than the bulk lattice constant of 4.177Å for NiO.[24] While, $a_\perp$lies below the bulk value. This suggests the existence of a tensile biaxial strain in these films. It should be noted that Yamauchi et al. gave a new concept of lattice matching based on a layer matching epitaxy (LME) model in case of (111) NiO layers grown on (0001) plane of sapphire[31], according to which the length mismatch between Ni (or O) equilateral triangle of (111) NiO plane and the equilateral triangle associated with oxygen sublattice of (0001) sapphire plane can be as low as 3% . However, this lattice mismatch should have introduced a compressive biaxial strain in the film. Observation of tensile strain in these layers may indicate that the thermal expansion coefficient mismatch between the film and the substrate must be the main factor behind this strain development. In fact, it should be noted that the thermal expansion coefficient of NiO is more than double the value of that of sapphire. Such a scenario can indeed explain in-plane tensile strain in the film, provided the layer is sufficiently thick (more than the critical thickness). In a thick epitaxial film, the lattice mismatch induced compressive strain is significantly suppressed by the generation of misfit dislocations.[32,33] In fact, the existence of misfit dislocations [see Fig. 3(c)] may evidence the relaxation of lattice mismatch induced strain in these layers.

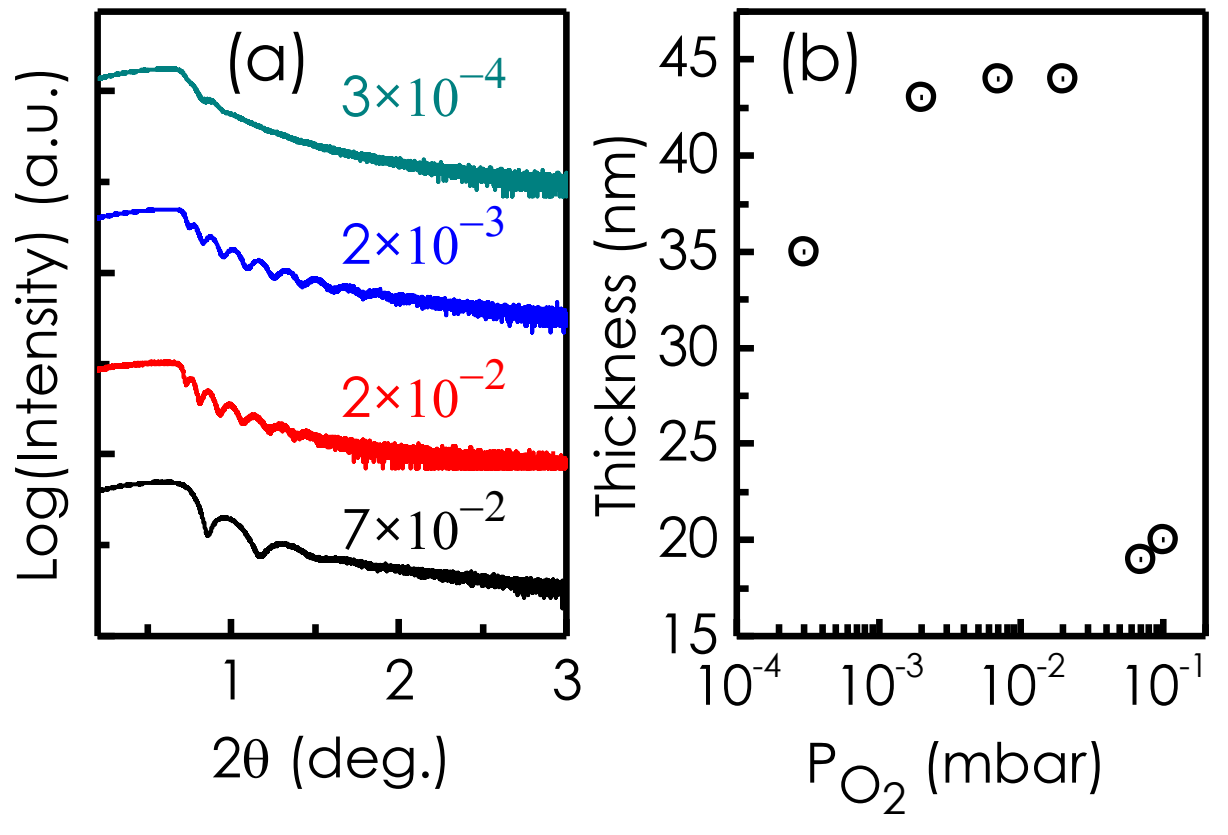

Fig.4 (a) XRR profiles for the samples grown at different $P_{O_2}$. (b) Layer thickness as a function of $P_{O_2}$. Error in thickness estimation is ~ $\pm 0.1$ nm.

Figure 4(a) shows the x-ray reflectivity (XRR) profiles for samples grown at various oxygen partial pressures. Clear intensity oscillations as a function of angle could be seen for these samples suggesting the growth of thin continuous films. Layer thickness obtained from XRR profiles is plotted as a function of $P_{O_2}$ in panel (b). Sudden fall of thickness for the two samples, which are grown with the highest oxygen pressures ($P_{O_2} > 2 \times 10^{-2}$ mbar), can be attributed to the reduction of the size of the plasma plume. At a high enough oxygen pressure, the plume size reduces, which keeps away adequate flux from reaching the substrate surface.

Figure 5(a) compares the absorption profiles for samples grown with $P_{O_2} = 1 \times 10^{-3}$ and $3 \times 10^{-4}$ mbar. Formation of tail states is quite evident (see the inset) for the sample grown with lower partial pressure of oxygen. These states can thus be attributed to oxygen deficiency related defects. Figure 5(b) plots the variation of the band gap ($E_g$) estimated from the absorption profiles as a function of $P_{O_2}$. Band gap has been found to range between 3.7 to 3.8 eV for all samples. However, it should be noted that the tail formation at the absorption edge for samples grown at low oxygen pressures make the correct estimation of band gap difficult.

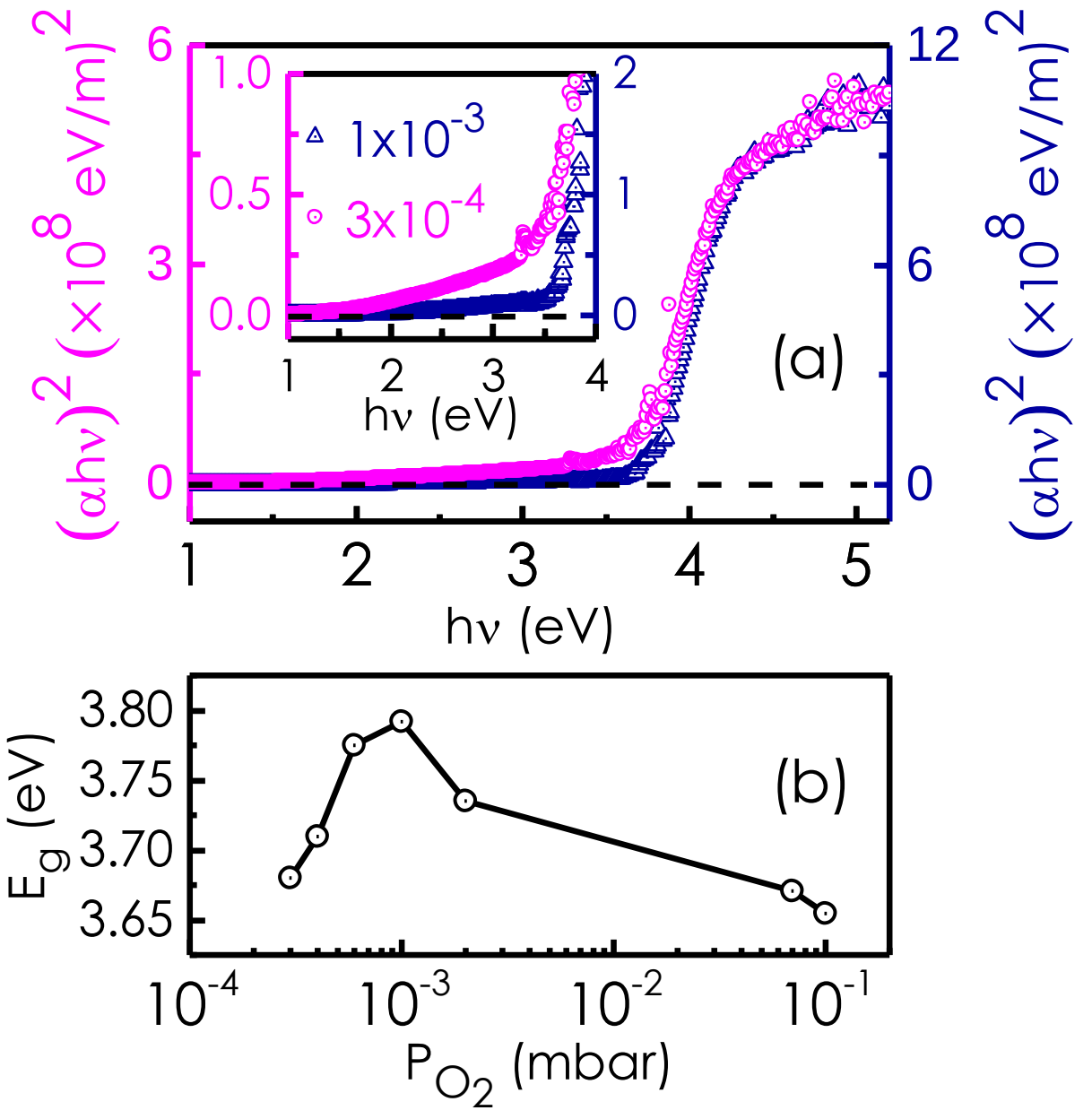

Fig. 5 Absorption profiles for samples grown with different oxygen partial pressures. Inset shows the profiles in an expanded scale at the on-set. (b) Band gap ($E_g$) as a function of $P_{O_2}$.

## B. Growth Temperature

This section deals with the films grown at various growth temperatures $T_G$ keeping oxygen partial pressure constant at $2 \times 10^{-2}$ mbar and the number of laser pulses at 10000. Thickness of these samples (as estimated by XRR study) is found to range from 37 to 47 nm. This series of samples is termed as $T_G$-series.

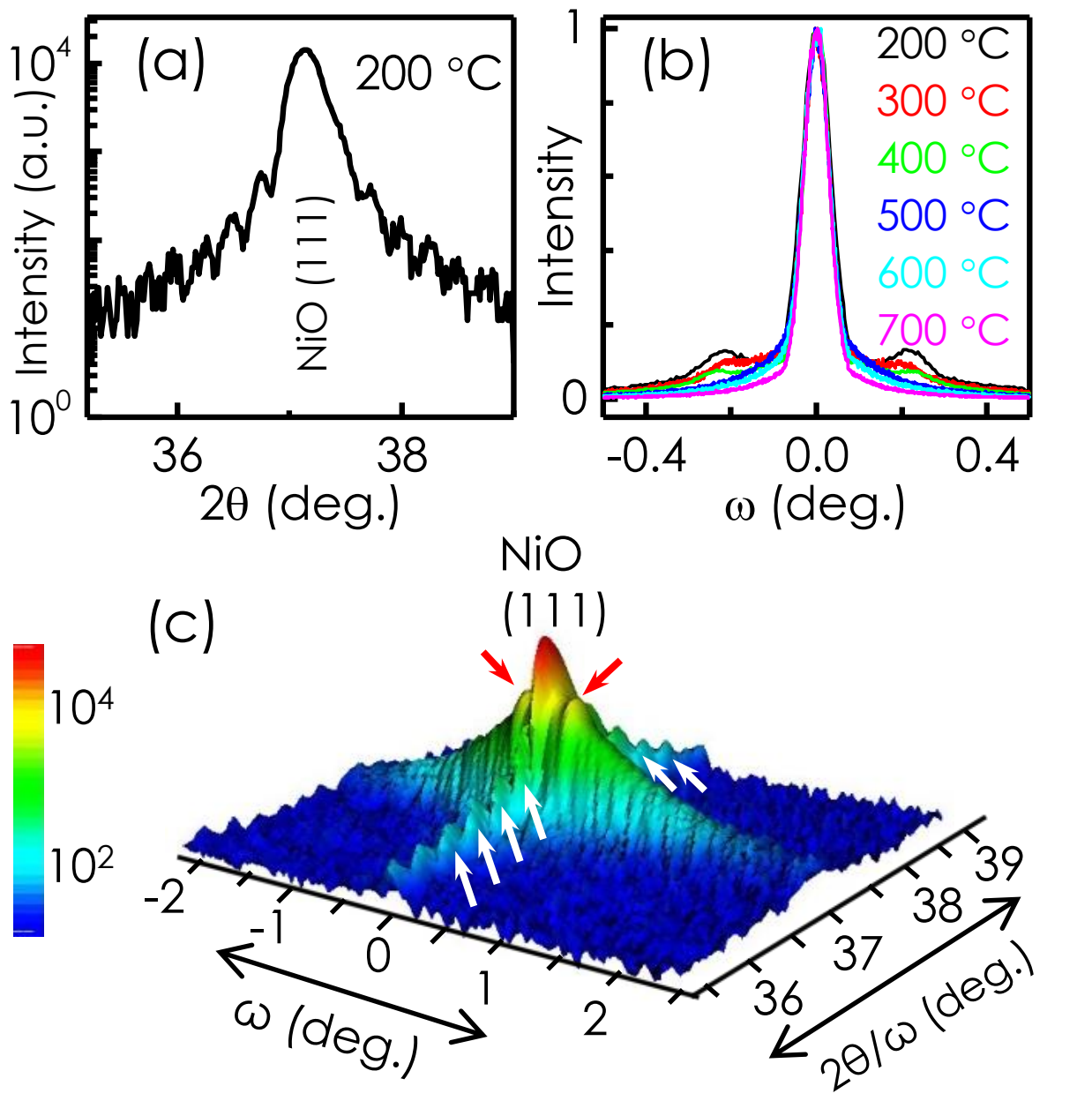

Fig. 6 (a) ω-2θ scan for the sample grown at $T_G$=200°C. (b) $\omega$-scans for the samples grown at different $T_G$. (c) Symmetric RSM for the sample grown at 200°C.

All samples of this series show only (111) and its higher order reflections in ω-2θ XRD scans, suggesting [111] directional growth of these films. Figure 6 (a) shows the (111) ω-2θ profile in an extended scale for the film grown at 200°C. Multiple satellites could be seen on both sides of the peak. These features, which are found almost in all the NiO samples investigated here, can be attributed to the interference effect. Appearance of these satellites is the mark of a superior quality interface between the film and the substrate. Figure 6(b) compares the (111)-NiO rocking curves ($\omega$-scan) recorded for the samples grown at different growth temperatures. Interestingly, for samples grown at lower growth temperatures, two satellite peaks are noticeable on either side of the main peak. Intensity of these peaks decreases as the growth temperature is enhanced. Beyond 400°C, these peaks can no longer be resolved. Instead, extended tails are formed at the two sides of the main peak, intensity of which gradually decreases as $T_G$ increases. It should be noted that the formation of extended tails at the two sides of the main peak has already been observed in case of $P_{O_2}$-series of samples [see Fig. 2(b)]. The two satellite features can be attributed to 60°-dislocations with opposite tilt components.[34] It has been shown by Kaganer et al. that such satellites can be observed in symmetric $\omega$-scans only when the density of these dislocations is very low in the film. Note that the intensity of the tail parts of the peak decreases with the increase of $T_G$. This implies a reduction of the density of 60°-dislocations with the increase of growth temperature in these films. It should be noted that in $P_{O_2}$-series of samples, tail broadening is found to increase with the reduction of $P_{O_2}$ suggesting an enhancement in the density of 60°-dislocations as $P_{O_2}$ decreases. Furthermore, like in the $P_{O_2}$-series of samples, the width of the main peak is also very narrow [FWHM ~0.08°] for these layers implying a low density of screw type dislocations.

Fig. 6(c) shows the topographic plot of the symmetric RSM for the sample grown at 200°C. Two satellite features are clearly visible (marked by red arrows) on either side of (111) NiO reflection in the $q_x$-direction ($\omega$-scan). While, along the $q_z$-axis ($2\theta/\omega$ scans), the intensity distribution is featured by multiple satellites (marked by white arrows). These satellites can be attributed to the finiteness of the layer thickness and their formation suggests the dominance of coherent diffraction over the diffused component arising due to dislocations.[34] In other words, the density of dislocations, which affects the symmetric diffraction of x-ray, must be very low in this sample.

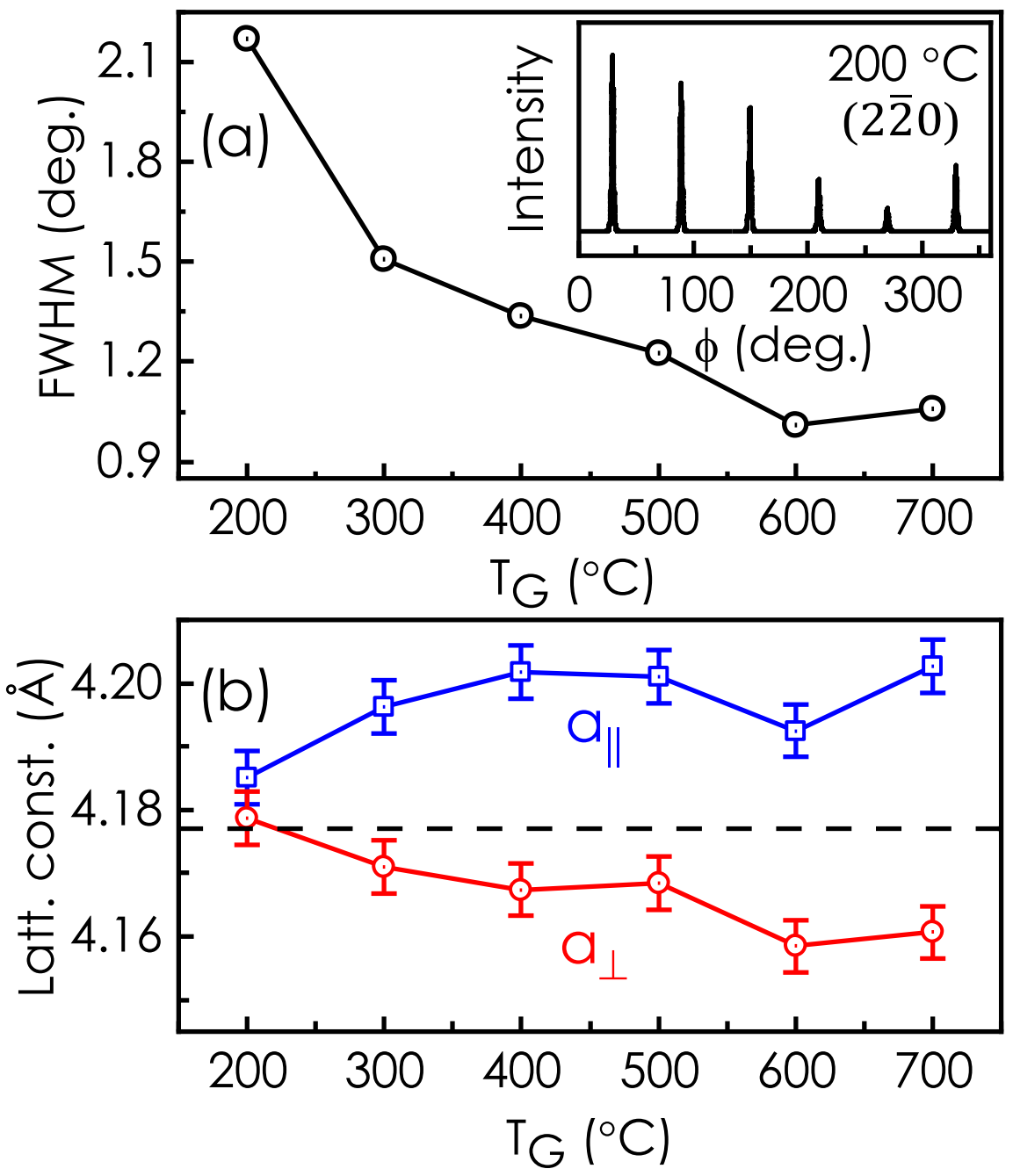

Fig. 7 (a) FWHM of the in-plane $\phi$-scans for the (2$\bar{2}$0) reflection as a function of $T_G$. Error in FWHM estimation is ~0.1%. Inset shows the in-plane $\phi$-scan for the sample grown at 200°C. (b) Lattice constants $a_{\parallel}$ and $a_{\perp}$ as a function of $T_G$.

Figure 7(a) plots the FWHM of the in-plane (2$\bar{2}$0) $\phi$-scans for the samples as a function of $T_G$. Evidently the width decreases with the growth temperature suggesting a reduction of density of edge dislocations with the increase of $T_G$. Note that the trend is similar to that of the 60° dislocations. Figure 7(b) plots the variation of the lattice parameters $a_{\parallel}$ and $a_{\perp}$ as a function of $T_G$. Presence of a tensile biaxial strain in these samples is quite evident. Clearly, the strain increases as the growth temperature increases, which might support our previous argument that the thermal expansion coefficient mismatch between the layer and the substrate is the main reason for this biaxial strain.

**C. Thickness Dependent**

Another set of samples is grown for different number of laser pulses ranging from 500 to 15000 keeping $T_G$ = 600°C and $P_{O_2}$ = 2× $10^{-2}$ mbar fixed.

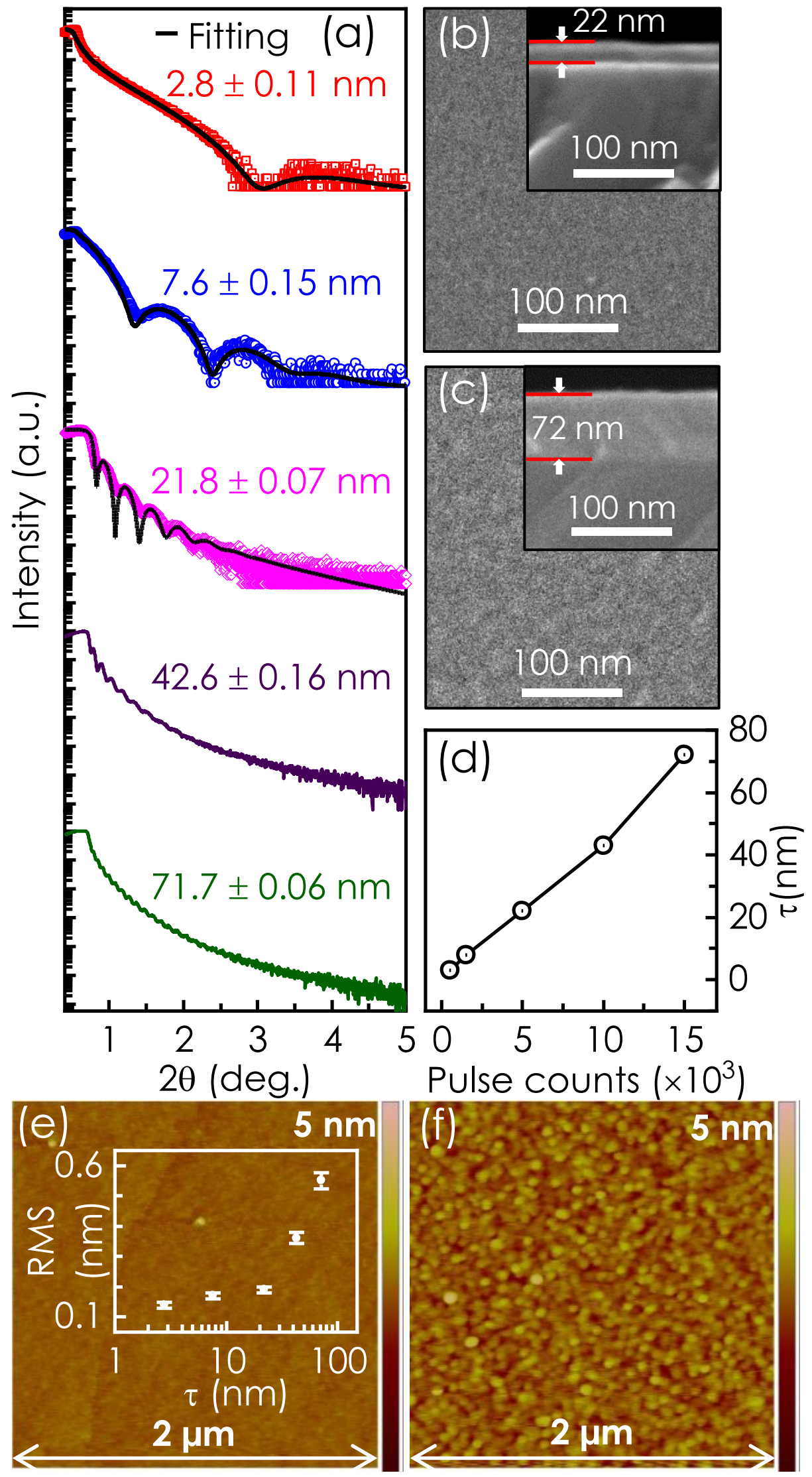

Fig. 8 (a) XRR profiles for the samples grown at different pulse counts keeping $P_{O_2}$ and $T_G$ constant. SEM surface and cross-sectional (inset) images for samples grown for pulse counts of (b) 5000 and (c) 15000. (d) Layer thickness as a function of pulse counts. AFM images for the films with thickness (e) 3 and (f) 43 nm. Inset of (e) shows the RSM surface roughness of films with various layer thicknesses.

Figure 8(a) shows the XRR profiles for the samples grown with various pulse counts. Fitting of the profiles for three samples are shown as examples. Fig. 8(b) and 8(c) show the SEM surface and cross-sectional (inset) images for two samples. Deposition of a smooth and continuous layer is quite evident for both the samples. Moreover, it is noteworthy that the layer thickness obtained through XRR analysis matches very well with that is found from cross-sectional SEM images for these films. Thickness ($\tau$) of these films shows a linear increase with the pulse number as can be found in Fig. 8(d). Fig. 8 (e), (f) present the AFM top-view images for the two samples. Clearly, the surface of the thicker layer is rougher than the thinnest one. RMS roughness of these films has indeed been found to increase with the thickness as shown in the inset of Fig. 8(e). More results on the AFM, SEM, XRR and XRD scans of these layers can be found in the supplementary information.

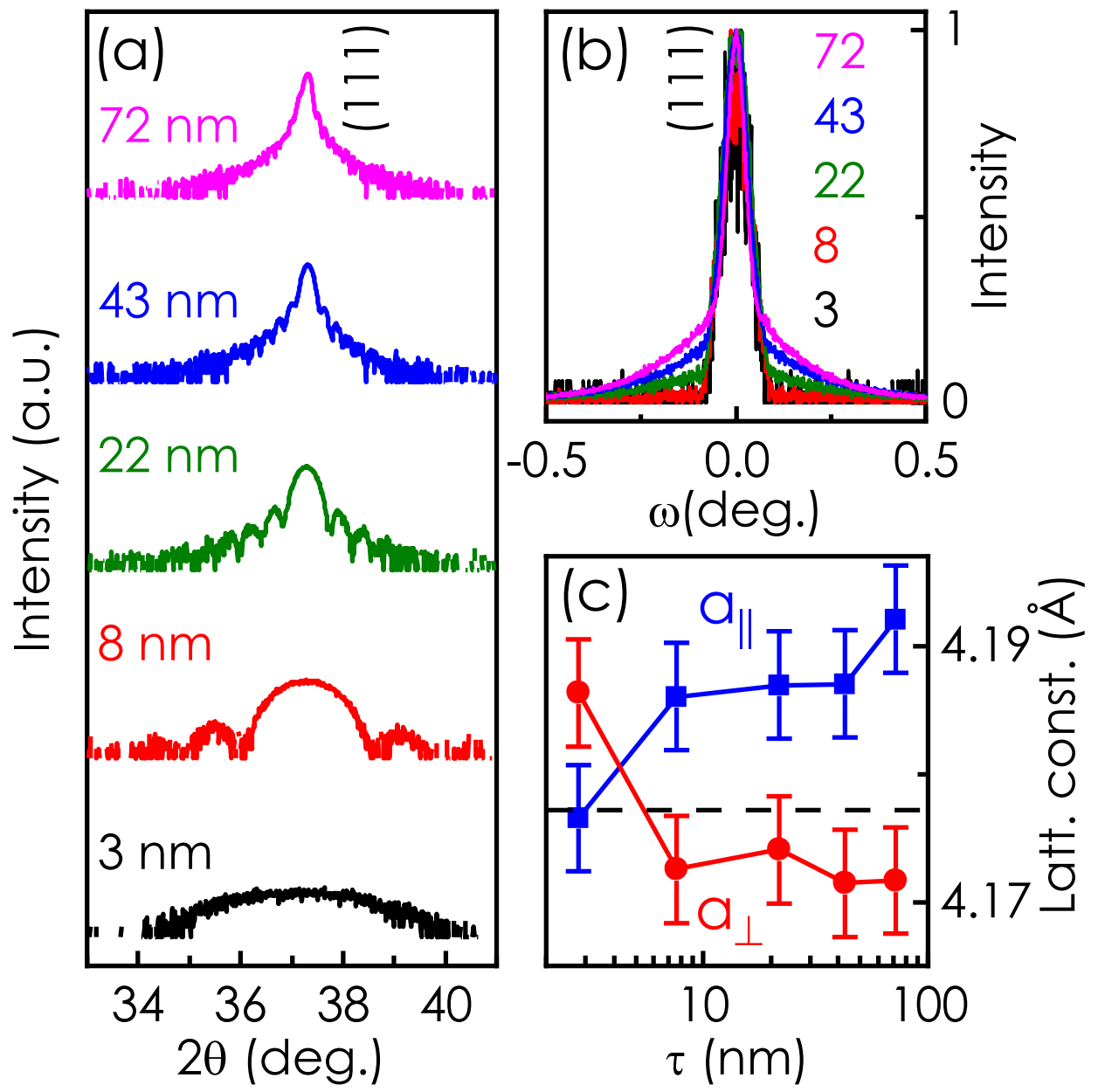

Fig. 9 (a) $\omega - 2\theta$ scans for samples grown at different layer thicknesses. (b) Rocking curves for (111)-NiO reflection. (c) Lattice constants $a_{\parallel}$ and $a_{\perp}$ as a function of $\tau$.

(111) and its higher order (222) reflection are the only features of NiO visible in ω-2θ XRD scans for these samples suggesting [111] directional growth. Figure 9(a) compares ω-2θ profiles around (111) reflection for these samples. Broadening of the feature with the reduction of layer thickness due to the finite size effect is clearly visible. Appearance of multiple satellites, which is also evident from the figure, implies a good layer-substrate interface quality in these films. In Fig. 9(b), normalized (111) $\omega$-scan profiles for these samples are compared. It is noteworthy that the width of the symmetric rocking curves remains to be quite narrow (~ 0.08°) and does not show any tangible variation with $\tau$. However, the broadening of the tail parts of the feature has been found to increase with the thickness. This may suggest an enhancement in density of 60°-dislocations with $\tau$, while the screw dislocation density does not change much. Fig. 9(c) shows the variation of $a_{\parallel}$ and $a_{\perp}$ lattice parameters as a function of $\tau$. Evidently, the thinnest sample shows a compressive biaxial strain, while the biaxial strain in rest of the samples is tensile in nature. As described earlier, the origin of the tensile strain for the thicker films is believed to be the mismatch in thermal expansion coefficients of the layer and the substrate. While, in the thinnest film, the layer-substrate lattice mismatch should play the main role in strain generation. Note that the lattice mismatch between (111) plane of NiO and c-plane of sapphire should lead to a compressive biaxial strain in NiO film.[10] This explains why the polarity of the strain in the thinnest sample is opposite to that for the other samples.

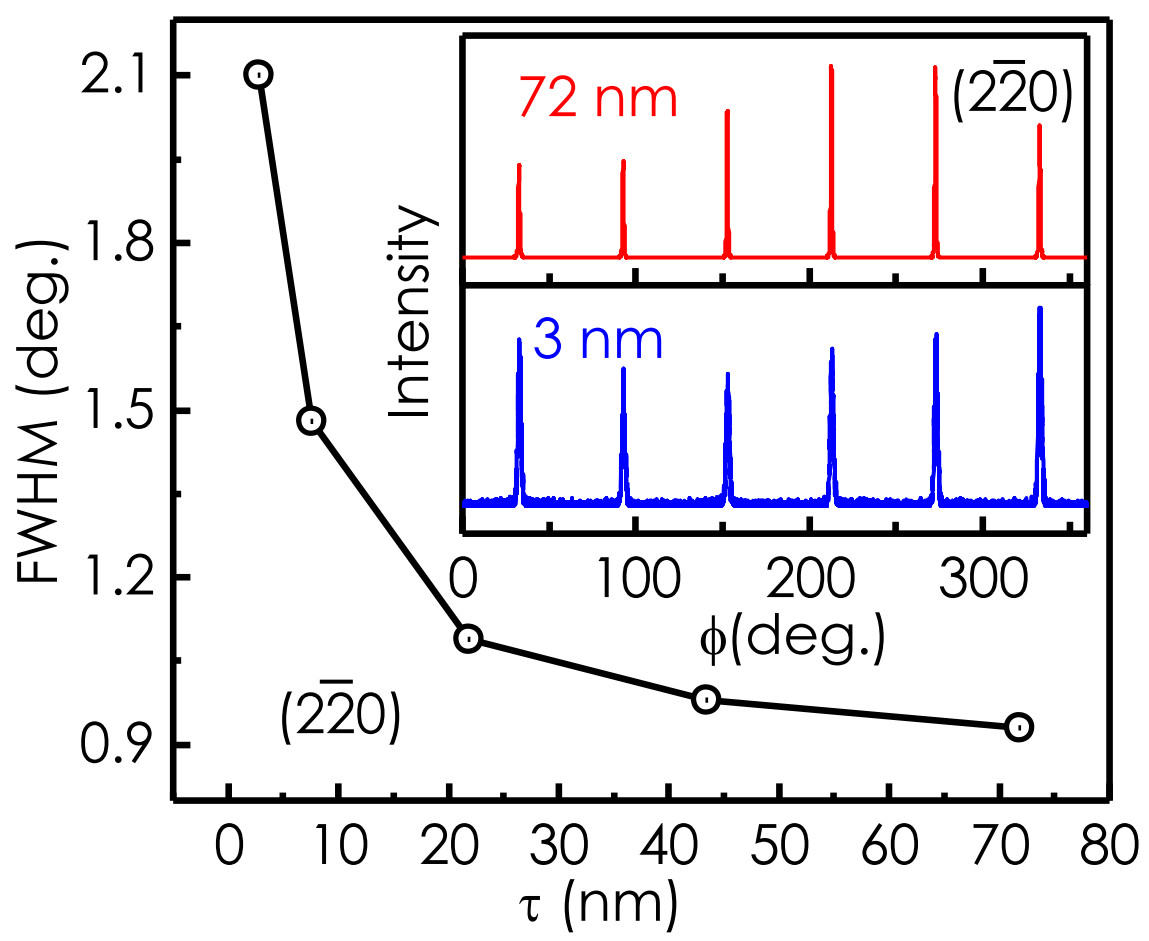

Fig.10 FWHM of the in-plane $\phi$-scans for the $(2\bar{2}0)$ reflection as a function of $\tau$. Error in FWHM estimation is ~0.1%. Inset compares the in-plane $\phi$-scans for the samples with different thicknesses.

Inset of Fig. 10 presents the $(2\bar{2}0)$ $\phi$-scans for the two samples. Observation of six equidistant peaks, which are also observed for other samples of the series, confirms the epitaxial nature of these films even when the layer thickness is only 3 nm. Fig 10 plots the FWHM for $(2\bar{2}0)$ $\phi$-scans as a function of the layer thickness ($\tau$). Evidently, the width decreases as $\tau$ increases showing a reduction of edge dislocation density with the increase of film thickness.

As mentioned before that the study shows a very narrow out-of-plane rocking curve width (FWHM ~0.08°), which corresponds to a screw dislocation density as low as $3 \times 10^8$ cm$^{-2}$ in these layers. Such a low value of broadening is the mark of a high crystalline quality of these films. Similarly, the broadening for in-plane rocking curves ($\phi$-scans) is also found to be less than 1° corresponding to the edge dislocation density less than 6.5 $\times 10^{10}$ cm$^{-2}$. This value is quite typical in high quality heteroepitaxial films. It should be noted that out-of-plane XRD studies on epitaxial NiO films are also carried out by other groups.[3,10,23] In a few cases the rocking curves show similarly narrow FWHM.[3,10] However, before this study, not much has been done to evaluate the in-plan orientation of the grains in NiO epitaxial films.

## IV. CONCLUSION

High quality epitaxial (111) NiO films can be grown on c-sapphire substrate by pulsed laser deposition (PLD) technique. Continuous epitaxial layers of thickness as low as 3 nm with high crystalline quality, smooth surface and interface morphology can be achieved by this method. It has been found that these films are consisting of 60°-rotated (111) NiO triangular domains. The density of screw and 60-degree dislocations in these samples is found to be very low. Interestingly, the density of screw dislocation does not vary much with the growth parameters. While, the 60-degree dislocations show an increasing trend with the layer thickness and a tendency of reduction with the increase of oxygen pressure and growth temperature. On the other hand, the edge dislocation density is found to decrease with the increase of growth temperature and layer thickness, but it increases with oxygen partial pressure. Barring the thinnest one, a tensile biaxial strain is present in all samples. This is likely to be resulting from the mismatch in thermal expansion coefficients between the layer and the substrate. A biaxial compressive strain is found in the thinnest layer. This can be attributed to the lattice mismatch between the layer and the substrate.

## ACKNOWLEDGEMENT

We acknowledge financial support provided by Department of Science and Technology (DST) under grant No: CRG/2018/001343, Government of India. We would like to thanks Centre of Excellence in Nanoelectronics (CEN), Industrial Research and Consultancy Centre (IRCC), and Sophisticated Analytical Instrument Facility (SAIF), IIT Bombay for using various facilities.